%% file: main.tex
\documentclass[journal]{vgtc}                     %

\onlineid{1516}

\vgtccategory{Research}

\vgtcpapertype{application/design study}

\title{Story Ribbons: Reimagining Storyline Visualizations with \texorpdfstring{\\}{} Large Language Models}

\author{Catherine Yeh, Tara Menon, Robin Singh Arya, Helen He, Moira Weigel, Fernanda Vi\'egas, and Martin Wattenberg}

\authorfooter{
  \item Authors are with Harvard University. Vi\'egas  and Wattenberg are also with Google, but this work was done at Harvard. Emails: \{catherineyeh, robinsingh\_arya, fernanda, wattenberg\}@g.harvard.edu, 
  \{taramenon, weigel\}@fas.harvard.edu, helen\_he@college.harvard.edu.
}

\input{sections/00-abstract}

\keywords{Narrative visualization, interactive literary analysis, large language models}

\input{sections/00-teaser}

\graphicspath{{figs/}{figures/}{pictures/}{images/}{./}} %

\usepackage{tabu}                      %
\usepackage{booktabs}                  %
\usepackage{lipsum}                    %
\usepackage{mwe}                       %

\usepackage{mathptmx}                  %

\usepackage{framed}
\usepackage{tabularx}
\usepackage{relsize}
\usepackage{xspace}
\usepackage{xpunctuate}
\usepackage{enumitem}
\setlist{nosep}
\usepackage{mdframed}

\usepackage{soul}
\setul{0.4ex}{0.2ex}

\begin{document}

\input{sections/00-macros}

\input{sections/01-intro}

\input{sections/02-related}

\input{sections/03-tasks}

\input{sections/04-pipeline}

\input{sections/05-design}

\input{sections/06-eval}

\input{sections/07-discussion}

\input{sections/08-conclusion}

\input{sections/00-acknowledgements}

\bibliographystyle{abbrv-doi-hyperref}

\bibliography{references}
\end{document}

%% file: sections/00-abstract.tex
\abstract{%
    Analyzing literature involves tracking interactions between characters, locations, and themes.
    Visualization has the potential to facilitate the mapping and analysis of these complex relationships, but capturing structured information from unstructured story data remains a challenge. 
    As large language models (LLMs) continue to advance, we see an opportunity to use their text processing and analysis capabilities to augment and reimagine existing storyline visualization techniques.
    Toward this goal, we introduce an LLM-driven data parsing pipeline that automatically extracts relevant narrative information from novels and scripts. 
    We then apply this pipeline to create \system, an interactive visualization system that helps novice and expert literary analysts explore detailed character and theme trajectories at multiple narrative levels. 
    Through pipeline evaluations and user studies with \system on \totalcount literary works, we demonstrate the potential of LLMs to streamline narrative visualization creation and reveal new insights about familiar stories. We also describe current limitations of AI-based systems, and interaction motifs designed to address these issues.
}

%% file: sections/00-teaser.tex
\teaser{
  \centering
  \includegraphics[alt={Screenshot of our \system interface, which shows a partial visualization of \textit{Pride and Prejudice.}},width=\linewidth]{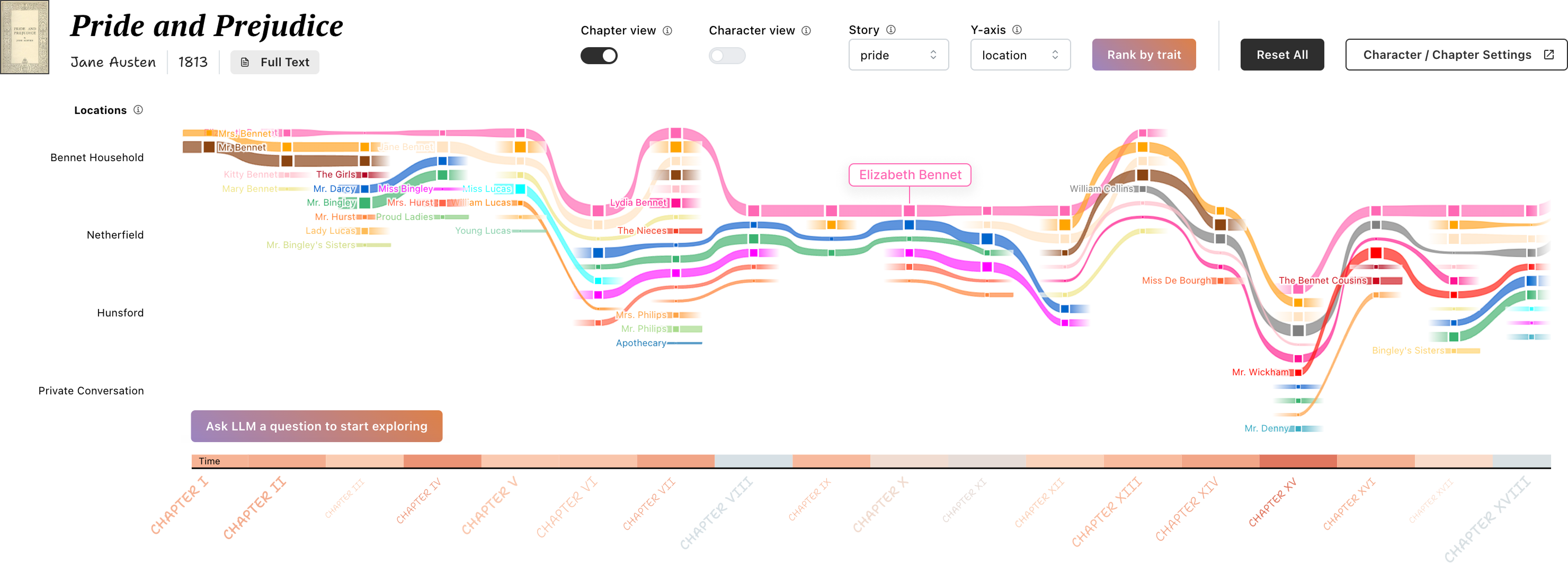}
  \caption{\system is an interactive narrative analysis tool that visualizes LLM-extracted insights about literary works. 
  Here, a partial visualization of \textit{Pride and Prejudice} by Jane Austen is shown. 
  Each ``ribbon'' represents a different character (\eg the top \textcolor{BrightPink}{pink} ribbon = Elizabeth Bennet), and can be used to track interactions across novel chapters (x-axis) and locations (y-axis).
  Chapter titles are colored by sentiment (\textcolor{DarkRed}{red}: positive, \textcolor{DarkBlue}{blue}: negative).
  \system enables users to explore stories at multiple narrative levels, and offers several features to customize visualizations to individual analysis workflows.
  }
  \label{fig:teaser}
}

%% file: sections/00-macros.tex
\newcommand{\ie}{{i.e.,}\xspace}
\newcommand{\eg}{{e.g.,}\xspace}
\newcommand{\ea}{{et~al\xperiod}\xspace}
\newcommand{\aka}{{a.k.a.}\xspace}
\newcommand{\etc}{{etc\xperiod}\xspace}
\newcommand{\etal}{{et al\xperiod}\xspace}
\newcommand{\vs}{{vs.}\xspace}

\newcommand{\todo}[1]{{\textcolor{OrangeRed}{[Todo: #1]}\normalfont}}
\newcommand{\cy}[1]{{\textcolor{RoyalBlue}{[CY: #1]}\normalfont}}

\newcommand{\system}{\textsc{Story Ribbons}\xspace} %
\newcommand{\totalcount}{36\xspace} %
\newcommand{\novelcount}{21\xspace} %
\newcommand{\playcount}{5\xspace} %
\newcommand{\poemcount}{2\xspace} %
\newcommand{\othercount}{2\xspace} %
\newcommand{\llmcount}{6\xspace} %
\newcommand{\numscholars}{3\xspace} %
\newcommand{\numparticipants}{16\xspace} %
\newcommand{\numexperts}{5\xspace} %
\newcommand{\numintermediate}{5\xspace} %
\newcommand{\numnovices}{6\xspace} %
\newcommand{\numguides}{6\xspace} %
\definecolor{BrightPink}{HTML}{DE408F}
\definecolor{MediumBlue}{HTML}{6598DB}
\definecolor{MediumPink}{HTML}{E56A8F}
\definecolor{MediumPurple}{HTML}{9473C8}
\definecolor{MediumOrange}{HTML}{ED782F}
\definecolor{MediumGreen}{HTML}{64B24F}
\definecolor{CoralRed}{HTML}{EA5246}
\definecolor{DarkRed}{HTML}{670000}
\definecolor{DarkBlue}{HTML}{08306B}
\definecolor{Gold}{HTML}{E6AE2C}
\definecolor{Teal}{HTML}{4399A8}
\definecolor{Magenta}{HTML}{9C679C}
\definecolor{storybg}{HTML}{F5F5F5}
\definecolor{DarkGray}{HTML}{666666}
\newcommand{\decompose}[1]{\textcolor{MediumBlue}{#1}}
\newcommand{\aggregate}[1]{\textcolor{MediumPink}{#1}}
\newcommand{\reflink}[2]{\hypersetup{hidelinks}\textcolor{Black}{\hyperlink{#1}{#2}}}
\newcommand{\point}{\includegraphics[width=0.25cm]{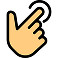}\xspace}
\newcommand{\robot}{\includegraphics[width=0.25cm]{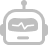}\xspace}
\newcommand{\light}{\includegraphics[width=0.25cm]{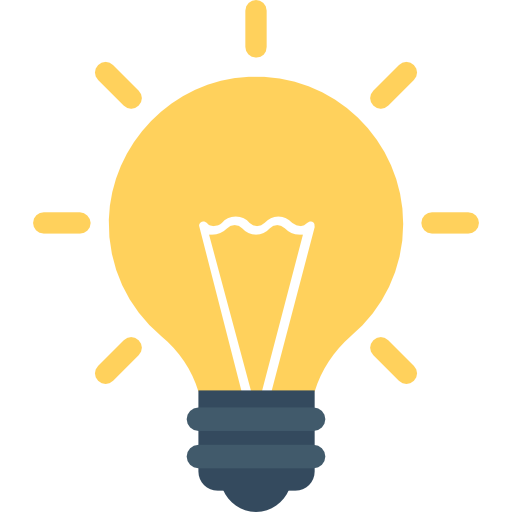}\xspace}
\newcommand{\book}{\includegraphics[width=0.25cm]{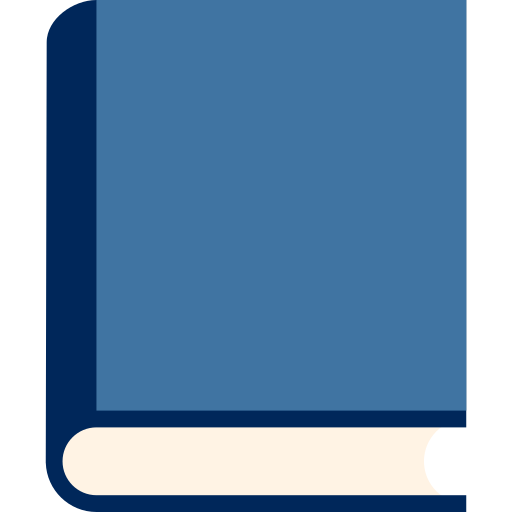}\xspace}
\newcommand{\girl}{\includegraphics[width=0.25cm]{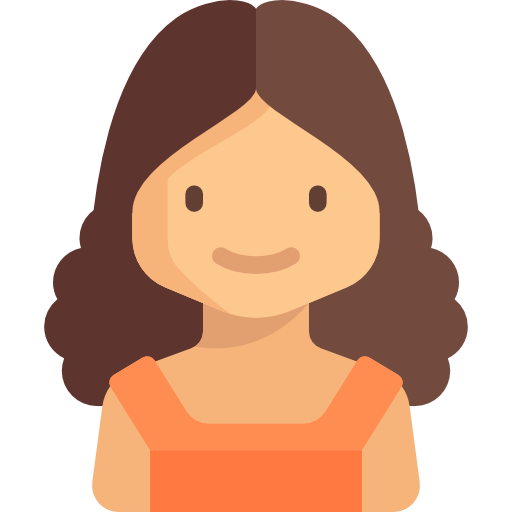}\xspace}
\newcommand{\house}{\includegraphics[width=0.25cm]{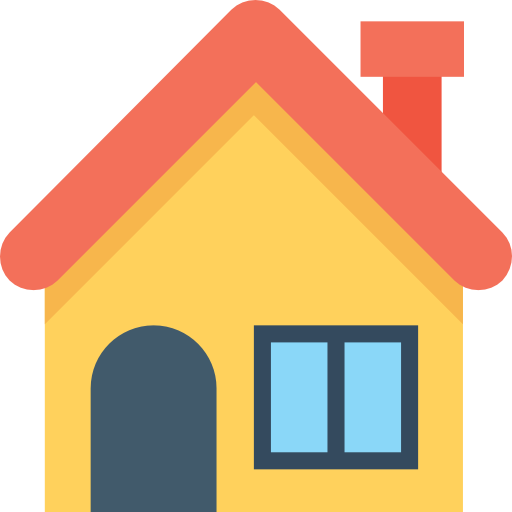}\xspace}
\newcommand{\theme}{\includegraphics[width=0.25cm]{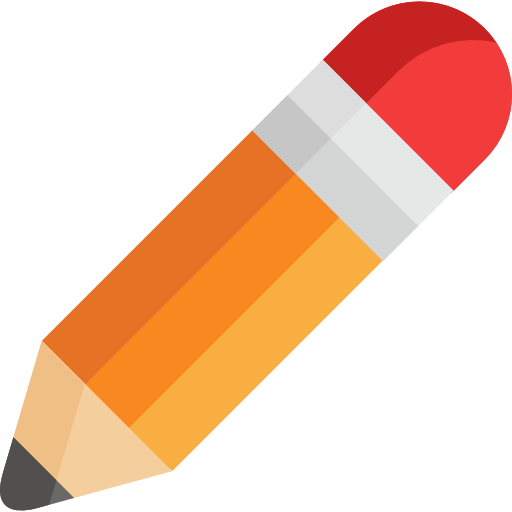}\xspace}
\newcommand{\chat}{\includegraphics[width=0.25cm]{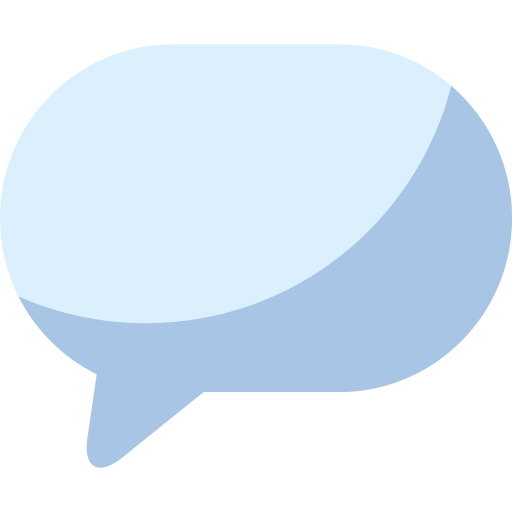}\xspace}
\newcommand{\repeaticon}{%
  \begingroup\normalfont
  \includegraphics[height=1.2\fontcharht\font`\B]{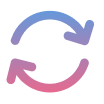}%
  \endgroup
}
\newcommand{\famous}{\includegraphics[width=0.15cm]{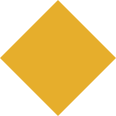}\xspace}
\newcommand{\lessknown}{\includegraphics[width=0.12cm]{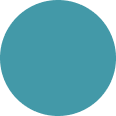}\xspace}
\newcommand{\llmstory}{\includegraphics[width=0.15cm]{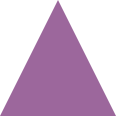}\xspace}
\newcommand{\pinkspark}{\includegraphics[width=0.2cm]{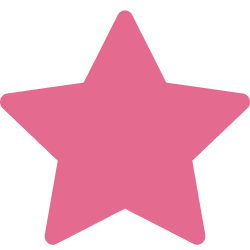}\xspace}
\newcommand{\purplespark}{\includegraphics[width=0.2cm]{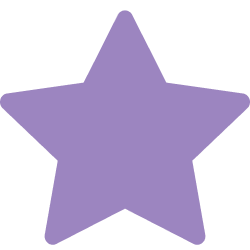}\xspace}
\newcommand{\orangespark}{\includegraphics[width=0.2cm]{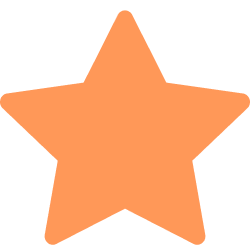}\xspace}
\newcommand{\explain}[1]{\purplespark\textcolor{MediumPurple}{\textit{#1}}}
\newcommand{\customize}[1]{\pinkspark\textcolor{MediumPink}{\textit{#1}}}
\newcommand{\explore}[1]{\orangespark\textcolor{MediumOrange}{\textit{#1}}}
\newcommand{\feedbackloopbox}[3]{
    \vspace{0.25em}
    \begin{minipage}{\linewidth} %
        \FrameSep7pt
        \begin{framed}
            \footnotesize
            \setlength\parindent{0pt}
            \textbf{\repeaticon\hspace{0.25em} Correction Loop: {#1}}
            \vspace{1pt}\\
            #2
            \vspace{3pt}\\
            \textit{\textbf{Solution:}} #3
        \end{framed}
    \end{minipage}
}

\newmdenv[
  linewidth=0pt,                 %
  innerleftmargin=7pt,
  innerrightmargin=7pt,
  innertopmargin=7pt,
  innerbottommargin=7pt,
  skipabove=0.5em,
  skipbelow=0.5em,
  backgroundcolor=storybg       %
]{storyframe}

\newcommand{\storybox}[1]{
    \begin{minipage}{\linewidth}
        \begin{storyframe}
            \setlength\parindent{0pt}
            #1
        \end{storyframe}
    \end{minipage}
}

%% file: sections/01-intro.tex
\firstsection{Introduction}\label{sec:intro}
\maketitle

Visualizing textual data is currently a major challenge. 
The key difficulty lies in extracting structured information from natural language. 
Typically, researchers use dedicated algorithms, ranging from counting words to make tag clouds~\cite{hassan2024improving,kaser2007tag} to elaborate statistical methods such as topic modeling~\cite{min2019modeling,chu2022topic}. 
Yet, even the most sophisticated, bespoke approaches often fail to capture important aspects of meaning. 

Given recent successes of AI systems based on large language models (LLMs)\footnote{Technically, for the AI systems we used, language modeling is just one step in their training process. However, for brevity, we refer to them as LLMs.}, it is natural to ask whether their power and generality can help us build better text visualizations. 
Of course, LLMs are not a magic pixie dust that we can just sprinkle on existing visualizations for great results. 
They bring their own challenges, producing output that can be unpredictable, mysterious, and even contain ``hallucinations''\cite{kim2024fables,lewis2020retrieval}.

In this work, we explore how to harness the power of LLMs while addressing their limitations. Our focus is on visualizing stories, specifically through \textit{storyline visualization} techniques. 
Storyline visualizations portray narrative timelines~\cite{house2023clover,kim2017visualizing,segel2010narrative,watson2019storyprint} with the goal of helping people critically examine and interpret works of literature. 
For example, an ideal storyline visualization of \textit{Pride and Prejudice} by Jane Austen might show Elizabeth's evolving dynamic with Mr. Darcy, her sister Jane's gentle romance with Mr. Bingley, and Lydia's reckless elopement with Wickham -- all unfolding across distinct settings from the grand estate of Pemberley to the regimented world of Longbourn.

The issue is how to convert the raw story text into concrete representations of a ``gentle romance'' or ``reckless elopement.'' 
Extracting relevant information to visualize how characters interact and how their relationships shift is no simple task~\cite{hoque2023portrayal,liu2013storyflow}.
As such, preparing the input data for narrative visualizations -- \eg a film script or novel -- often requires extensive time and manual effort~\cite{kim2017visualizing}. 
This is especially challenging for novels, which do not contain metadata such as explicit scene divisions or character/location labels~\cite{hoque2023portrayal}.

LLMs seem like a promising approach for extracting the data needed for storyline visualizations~\cite{hoque2023portrayal,house2023clover,jaipersaud2024show,piper2024social,ye2024storyexplorer}.
Our work investigates how LLMs can help build and extend traditional storyline visualizations to aid literature analysis.
Concretely, our research questions \hypertarget{rq1}{are:} 

\begin{itemize}
    \item \textbf{RQ1:} How can we use LLMs to automate and extend the extraction of data for unstructured \hypertarget{rq2}{narratives?}
    \item \textbf{RQ2:} What are the right forms of visualization and interaction to help people understand and calibrate trust in AI story insights?
\end{itemize}

To address these questions, we created an interactive system, \system. 
Over six months, we co-designed the tool~\cite{burkett2012introduction,mccurdy2015poemage} with three literary scholars, who became co-authors on this paper (\textbf{C1-3}) and have expertise in narrative theory, comparative literature, and literary criticism.
\system lets users explore stories at multiple narrative levels, visualizing locations, characters, and themes along a rich and customizable set of literary dimensions (Fig.~\ref{fig:teaser}). 
The system is based on an LLM-powered data processing pipeline, which is almost completely automated, for extracting detailed narrative information from stories.

We then evaluated our system in multiple ways. 
First, we quantitatively and qualitatively assessed the data pipeline performance on \totalcount stories.
To evaluate our visualizations, we conducted a user study with \numparticipants participants with varying levels of literature expertise, asking each to explore a story of their choice using \system. 
Finally, we interviewed three additional literary scholars for expert feedback.

Our findings suggest that despite limitations, LLMs can meaningfully augment traditional text visualizations. 
Although LLMs proved unreliable at extracting information when used naively, we were able to design a data pipeline that was sufficiently reliable to be helpful to users. 
The flexibility of LLM-powered analysis allowed us to visualize a variety of high-level concepts, and enabled users to invent their own dimensions for visualization.
Furthermore, the fact that LLMs could provide justifications for their outputs helped in calibrating trust.

To summarize, our main contributions are:
\begin{itemize}
    \item \textbf{An LLM-powered pipeline} for extracting and organizing character, location, theme, and scene data from unstructured text. 
    We believe our design can be helpful to others working with LLMs. 
    \item \textbf{\system, an interactive literary analysis tool}. 
    The system illustrates what we believe are important LLM-based interaction motifs: providing custom text analytics on demand, as well as explanations for LLM-extracted information.
    \item \textbf{User study findings and expert feedback} highlighting how users interact with LLM-enhanced visualizations; namely, ways in which our tool is useful as well as areas for future research.
\end{itemize}

%% file: sections/02-related.tex
\section{Related Work}\label{sec:related}

The history of finding a visual form for a story is long and rich~\cite{dobson2011interactive,freytag1895technique,bernstein1998patterns}.
Our work centers on storyline visualization, popularized in 2009 by Randall Munroe's hand-drawn charts on \texttt{xkcd}~\cite{munroe2009xkcd}. 
While many early computational efforts to visualize storylines focused on optimizing  layouts~\cite{arendt2017matters,di2020storyline,liu2013storyflow,padia2019system,tanahashi2012design,tang2020plotthread,tang2018istoryline}, we aim to enrich these visualizations from a data and interaction perspective by leveraging novel LLM technologies. 
Additionally, in contrast to recent efforts on AI for automatic visualization generation~\cite{dibia2023lida,vaithilingam2024dynavis,wu2021ai4vis,yang2024foundation,narechania2020nl4dv}, we use AI to extract meaningful insights from stories to visualize.
Below, we outline current challenges and opportunities for creating narrative visualizations (Sec.~\ref{sec:related_vis}) and using NLP techniques to augment literary analysis (Sec.~\ref{sec:related_nlp}).

\subsection{Visualizing Narratives: Challenges \& Opportunities}\label{sec:related_vis}
Storyline visualizations help users analyze complex narratives across various domains ~\cite{di2020s,hoque2023portrayal,schwan2019narrelations,segel2010narrative,min2019modeling}, including news stories~\cite{costa2023newslines}, political relationship data~\cite{hulstein2022geo,pena2022hyperstorylines}, and interactions between LLM agents~\cite{lu2024agentlens}.
In literature and film, which is our focus, researchers have explored variations of traditional storyline visualizations~\cite{munroe2009xkcd}, such as hierarchical and radial layouts (\eg StoryPrint~\cite{watson2019storyprint} and StoryCake~\cite{qiang2016storycake}) or adding two time axes for nonlinear narratives (\eg Story Curves~\cite{kim2017visualizing}).

However, storyline visualizations are often limited in scalability and complexity due to the challenges of processing text data~\cite{van2009mapping}.
As described in~\cite{kim2017visualizing}: \textit{``To extract story elements (scenes, characters, etc.) we implemented a parser for segmenting a [movie] script... Unfortunately, not all scripts are well formatted... To work around this problem, we developed a tagging interface to fix the labels.''}
With novels, parsing is even more difficult, due to the lack of metadata such as explicit scene divisions and character labels~\cite{hoque2023portrayal}.
We aim to reduce the manual effort involved in processing unstructured stories, while maintaining data quality and faithfulness, by experimenting with LLM capabilities. 

\subsection{NLP-Enhanced Story Analysis}\label{sec:related_nlp}
Historically, computational forms of literary analysis have been fairly limited to vocabulary or syntactic measures, such as tracking word frequencies and average word lengths~\cite{keim2007literature}, analyzing concordances~\cite{scrivner2017interactive}, or exploring dependency links~\cite{van2009mapping}. 
Thus, with recent advances in natural language processing (NLP), researchers have begun to experiment with new analytical approaches.
For instance, to create Portrayal~\cite{hoque2023portrayal}, an interactive visualization system for character analysis, the authors developed an NLP pipeline to extract character traits from fiction novels using SoTA co-reference and sentiment analysis models. 
However, this process still required several elements of manual parsing and tagging, which is where we see an opportunity for LLMs to step in.

Given their impressive text processing and analysis capabilities, many works explore different ways of using LLMs to analyze stories~\cite{kim2024fables,shen2024heart}. 
In~\cite{piper2024social}, the authors train a small language model to understand literary social networks.
\cite{jaipersaud2024show} introduces a framework for prompting LLMs to uncover implicit character portrayals, and~\cite{piper2024using} studies the application of LLMs in narrative discourse understanding.
Most similar to our vision, StoryExplorer~\cite{ye2024storyexplorer} and Clover Connections~\cite{house2023clover} create LLM-enhanced, visualization-based interfaces to enhance user understanding of stories.
However, StoryExplorer is a human-in-the-loop system that requires user annotations.
Clover Connections uses LLMs to extract character traits, but we design and validate an almost fully-automated, LLM-driven data processing pipeline.
We also place a larger focus on calibrating user trust in AI-extracted literary insights.

%% file: sections/03-tasks.tex
\begin{figure*}
    \centering
    \includegraphics[alt={Diagram of our story analysis pipeline, with decomposition steps on the left (splitting text into chapters, splitting chapters into scenes), and aggregation steps on the right (generating summaries, outputting final story data).},width=\linewidth]{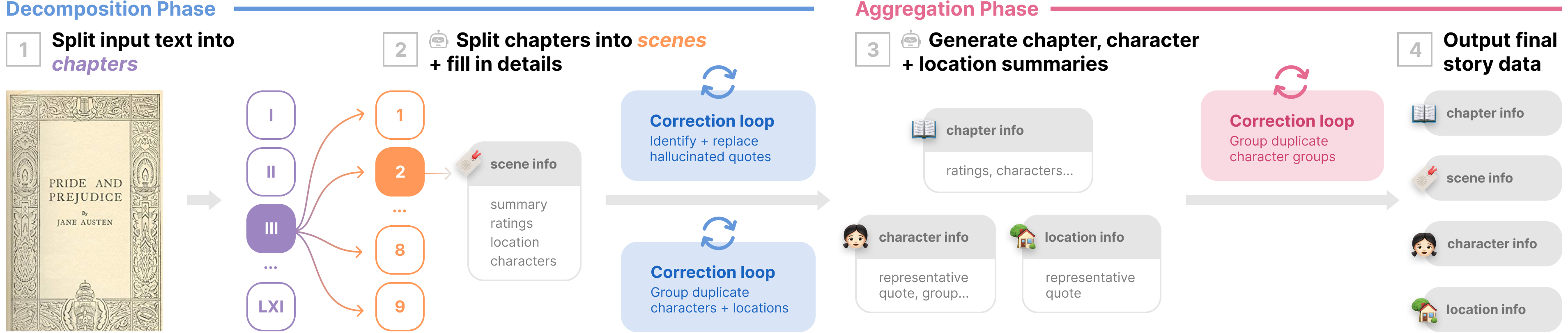}
    \caption{Overview of our story analysis pipeline, which is organized into a data decomposition and aggregation phase.
    Steps involving an LLM are denoted with \robot. 
    Correction loops are included to check and correct LLM output; each runs once per story.
    Our pipeline is highly customizable to different literary genres (\eg novels, plays) and elements --
    all steps involving \textit{character} data can be swapped out for another aspect such as \textit{theme}.}
    \label{fig:pipeline}
\end{figure*}
\section{Goals \& Tasks}\label{sec:tasks}
Literary analysis differs from conventional data analysis due to its open-ended, interpretive form. 
In contrast to analytical tasks where visualization is designed to uncover facts and numerical patterns, it often involves navigating multiple valid readings rather than converging on a single truth~\cite{mccurdy2015poemage,oelke2012advanced,house2023clover}. 
To explore how LLM-enhanced visualizations might facilitate this process, we interviewed three literary scholars, who are expert analysts: an English professor, a comparative literature professor, and an English Ph.D. student. 
We asked scholars (\textbf{C1-3}) about their current practices, as well as their hopes and concerns for incorporating LLMs into literary workflows.
Following~\cite{mccurdy2015poemage} and the tradition of co-design~\cite{burkett2012introduction,zamenopoulos2018co,steen2013co}\footnote{A collaborative research practice where real users of a system are included in the design process.}, these experts provided feedback throughout our design process and are co-authors of this paper.

\subsection{Design Goals}
Overall, scholars wanted to see how LLMs could enhance and extend their existing knowledge of literary works: \textit{``I feel like [LLM] analyses can be powerful in helping us see literature under a new interpretation''} (C2). 
Similarly, C3 viewed the prospect of integrating an LLM into their analysis process as \textit{``having a partner to bounce ideas off of, which can help you clarify your own perspective.''} 
C1 was curious if LLMs could help capture and visualize unexpected story patterns, as in~\cite{oelke2012advanced}: \textit{``I want to see surprising things in the visualization. For example, is one character a lot more prominent than others?''} 

From these formative discussions, we identified the following design goals to help analysts uncover new literary insights with \hypertarget{g1}{LLMs:}

\noindent \textbf{G1 - Support flexible analysis workflows.} Each scholar had unique analytical interests, reflecting the highly personalized nature of studying literature.
C1 noted that elements like \textit{``locations and themes [are] interesting, but not the most important [as] my work focuses on character prominence.''}
C2 was interested in \textit{``how characters are defined by language and dialogue,''} but said \textit{``settings are interesting''} as well. 

Scholars also performed different kinds of literary comparisons.
Some made absolute comparisons, \eg identifying the most important character in a scene (C2) or analyzing character gender distributions (C1). Others focused on dynamic trends, \eg how character networks and emotions change over time (C1, C3).
Given these diverse goals, we wanted our visualizations to be customizable to individual \hypertarget{g2}{users}~\cite{scrivner2017interactive}.

\noindent \textbf{G2 - Calibrate user trust and provide system transparency.}
Although scholars expressed optimism and curiosity about LLM-mediated literary analysis, they also shared concerns.
C3 explained the significant resistance in infusing technology into literary research: \textit{``Many scholars are still wary of AI, [and] think of it as something that might replace them,''} echoing~\cite{mccurdy2015poemage}.
C1 and C2 were also worried about LLM hallucinations, emphasizing that \textit{``it's important to point out when things might be hallucinated''} and provide clear explanations \textit{``in order to avoid making people feel suspicious.''}
Our work aims to prioritize trust and transparency, while limiting the potential for hallucinations.

\subsection{Design Tasks}
We then translated these goals into the following set of design \hypertarget{t1}{tasks:}

\noindent \textbf{T1 - Enable multiple levels of story exploration.}
To accommodate a wide range of analysis workflows \textbf{[\reflink{g1}{G1}]}, we support both \textit{high-level} (\eg chapters) and \textit{low-level} (\eg scenes) exploration of \hypertarget{t2}{stories.} 

\noindent \textbf{T2 - Track key story elements over time.}
Similarly, we visualize information about many story aspects such as \textit{characters}, \textit{locations}, and \textit{themes}, facilitating discovery based on user \hypertarget{t3}{interests} \textbf{[\reflink{g1}{G1}]}.

\noindent \textbf{T3 - Provide custom views on demand.}
Users can add new views by leveraging LLM capabilities, directly tailoring the visualization to answer individualized and spontaneous research \hypertarget{t4}{questions} \textbf{[\reflink{g1}{G1}]}.

\noindent \textbf{T4 - Explain AI decisions.}
We aim to provide explanations of LLM-generated data to maintain transparency and protect user \hypertarget{t5}{trust} \textbf{[\reflink{g2}{G2}]}.

\noindent \textbf{T5 - Connect visuals to raw text.}
To further calibrate trust \textbf{[\reflink{g2}{G2}]}, we link visualizations to the story text
so users can inspect AI insights.

%% file: sections/04-pipeline.tex
\section{Story Analysis Pipeline}\label{sec:pipeline}
To explore \textbf{\reflink{rq1}{RQ1}}, we created an LLM-powered analysis pipeline that automatically extracts and organizes narrative data from literary works. 

\subsection{Pipeline Design}\label{sec:pipeline_design}
We followed an iterative process to build our story analysis pipeline.
Our pipeline consists of an LLM chaining workflow that decomposes narrative processing into subtasks, inspired by common crowd programming patterns~\cite{grunde2023designing}.
The open-ended nature of literary analysis led to various undesirable model behaviors -- including hallucinations and inconsistent scene segmentations -- that required careful prompting and validation strategies.
For example, we implemented multiple \textbf{correction loops} throughout the pipeline to detect and correct unexpected LLM outputs, similar to quality checks used in crowdsourcing~\cite{daniel2018quality}.

\subsubsection{Overview}
Our pipeline contains four steps, comprising a decomposition (\decompose{\textbf{D}}) and aggregation (\aggregate{\textbf{A}}) phase~\cite{bernstein2010soylent} (Fig.~\ref{fig:pipeline}). 
We describe the process for analyzing \textit{characters} in a \textit{novel}, but our pipeline works for multiple genres (\eg plays, poems) and analysis targets (\eg themes).
Each correction loop runs once per story; see Sec.~\ref{sec:sys_implement} for full implementation details.

\noindent \textbf{1 - Split input text into chapters (\decompose{D}).} 
We first split the selected story (\eg \textit{Pride and Prejudice}) into \textit{chapters} (or \textit{acts} for a play).
\textbf{This is the only step requiring human assistance.}\footnote{This surprised us: identifying chapter boundaries does not seem hard. One reason may be that it requires the longest LLM context window.}
We tried using an LLM to identify chapter markers, \eg Chapter I, but the model often produced inaccurate results (\eg missing chapters or punctuation errors)~\cite{lewis2020retrieval}.

Each chapter is further split into lines and annotated with line numbers to help with text parsing.
These preprocessing steps were motivated by our observations that LLMs struggle with long contexts (\eg analyzing the entire story at once), leading to hallucinations, narrative chronology mistakes, and omissions of key events~\cite{kim2024fables,house2023clover}.

\noindent \textbf{2 - Split chapters into scenes and fill in details (\decompose{D}).} Next, we prompt the LLM to split each chapter into \textit{scenes} to extract key plot points from the story.
Initially, we saw inconsistent results across runs, suggesting that LLMs do not inherently have a clear sense of what a ``scene'' is~\cite{zehe2021detecting}.
C1 confirmed that scenes are a complex concept and there is not necessarily a ``ground truth,'' especially for novels, as \textit{``much of the language of scenes comes from plays and films... even with people, how you define a scene could depend on what you're looking at.''}

Ultimately, our literary scholars agreed that defining scenes based on changes in story location felt most sensible, which we implemented in our final pipeline. 
We found that providing this explicit definition to the LLM and asking the model to explain why it started a new scene, similar to chain-of-thought prompting~\cite{wei2022chain}, enhanced output consistency (to a degree, see Sec.~\ref{sec:scene_analysis}).

For each scene, we ask for a summary, the location, and ratings important to understanding a narrative (conflict~\cite{ware2010modeling,frermann2023conflicts}, importance~\cite{dobson2011interactive,otake2020modeling,zhang2021salience}, sentiment~\cite{elkins2022shapes,rebora2023sentiment,min2019modeling}).
Conflict and importance are specified between 0 and 1 (very high conflict or importance), while sentiment is rated between -1 (very negative) and 1 (very positive).\footnote{We use LLMs instead of task-specific models (\eg for sentiment analysis~\cite{chu2022topic}), as our goal was to explore the capabilities of LLMs for story analysis.}
The LLM also extracts \textit{characters} (or \textit{themes}) in this scene.
For each character, the LLM describes their sentiment~\cite{house2023clover,watson2019storyprint} and emotion~\cite{shen2024heart,hoque2023portrayal} (\eg ``excited and carefree''), finding a direct quote from the text as evidence.

Once all scene details are generated, we run two correction loops:

\noindent \feedbackloopbox{Check for Hallucinations}
{When extracting quotes, the LLM sometimes hallucinates or modifies story dialogue (\eg changing a third-person POV to first-person).} 
{We add an exact string match check, replacing all false or modified quotes with a brief LLM explanation of the character's emotions.}

\noindent \feedbackloopbox{Group Duplicate Elements}
{Characters and locations may be referred to by different names throughout the story (\eg Jane vs. Jane Bennet vs. Miss Bennet), which the LLM frequently fails to recognize on the first pass.}
{We use a second LLM to group duplicate elements to create the finalized character and location lists.}
    
\noindent \textbf{3 - Generate chapter, character, and location summaries (\aggregate{A}).} With the extracted scene details, we then compose:

\begin{itemize}
    \item \textit{Chapter summaries}, which contain a brief summary of each chapter, importance and conflict ratings,
    and a list of character and location counts. 
    For each unique pair of interacting characters, we ask the LLM to summarize their chapter interactions.
    \item \textit{Character summaries}, which contain a quote about each character and semantic group decided by the LLM (\eg ``main characters'').
    Each character is also assigned a unique color and explanation. 
    \item \textit{Location summaries}, which contain a quote about each location.
\end{itemize}

For character summaries, we run one more correction loop:

\noindent \feedbackloopbox{Group Duplicate Elements}
{As with character names, the LLM may create similar character groups (\eg Bennet family vs. family members).}
{We use a second LLM to group duplicate elements to create the finalized list of character groups.}

\noindent \textbf{4 - Output final story data (\aggregate{A}).} We output all structured \textit{chapter}, \textit{scene}, \textit{character}, and \textit{location} data as a single JSON file.

\subsection{Pipeline Evaluation}\label{sec:pipeline_evaluation}

\textbf{Data.} We assessed our pipeline on \totalcount stories, including 30 literary works from \href{https://www.gutenberg.org/}{Project Gutenberg} (\novelcount novels, \playcount plays, \poemcount poems, \othercount non-fiction; Tab.~\ref{tab:stories}).
To examine potential training data effects (\eg LLM memorization of popular texts), we tested both \lessknown \textit{lesser-known} ($n=8$) and \famous \textit{well-known} ($n=22$) stories.
For evaluation, we considered a story ``well-known'' if it has a \href{https://www.sparknotes.com/}{SparkNotes} \textit{and} \href{http://litcharts.com/}{LitCharts} study guide.
Story lengths also varied (mean: 8846 lines). 
Our shortest text is \textit{The Metamorphosis} (1752 lines) and longest is \textit{Ulysses} (25435 lines).

To further control for training data effects, the last 6 stories are synthetic novels authored by \texttt{gpt-4o-mini}.
We generated these novels using an outline-conditioned AI writing workflow~\cite{rashkin2020plotmachines} with similar iterative decomposition and synthesis steps as our data pipeline~\cite{bernstein2010soylent,grunde2023designing}.
These LLM-generated stories are shorter (mean: 1588 lines) but were unlikely to have appeared verbatim in the training data.

\subsubsection{Overall Performance}\label{sec:pipeline_performance}
\begin{figure}[t]
    \centering
    \includegraphics[alt={(A) Two vertical bar charts showing pipeline output statistics across LLM-generated stories, non-play texts, and plays, as well as character vs. theme analyses. (B) A horizontal stacked bar chart showing scene boundary types across LLM-generated stories, non-play texts, and plays.},width=\linewidth]{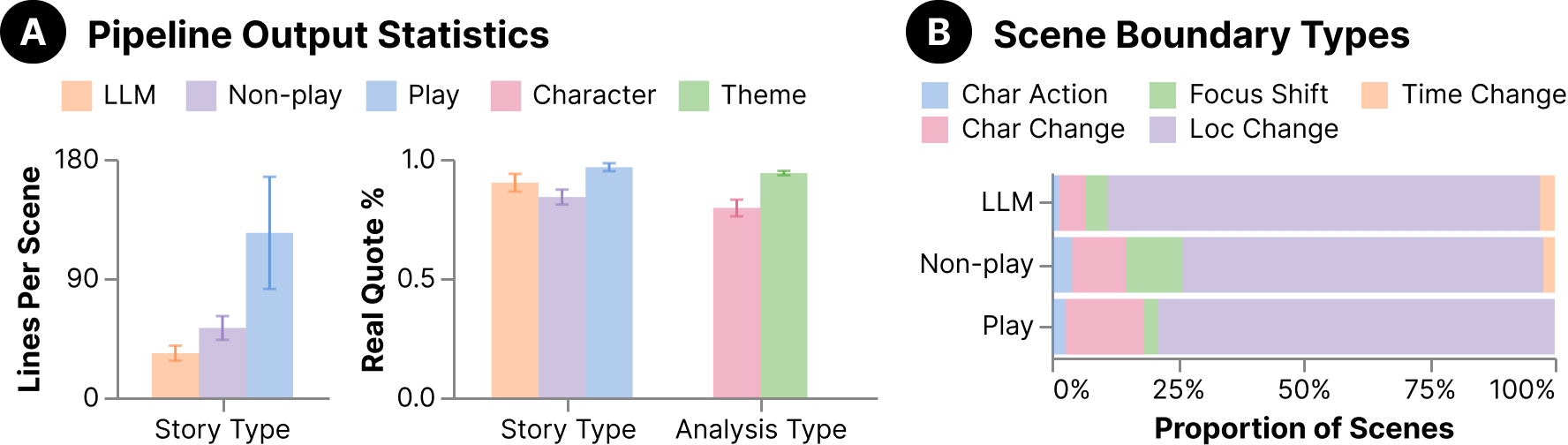}
    \caption{\textbf{(A)} Comparing the mean number of lines extracted per scene (\textit{left}) and percentage of real quotes (\textit{right}) identified across different story \& analysis types. 
    Error bars indicate 95\% CIs. \textbf{(B)} LLM classifications of scene divisions ($n=3796$) by story type. 
    }
    \label{fig:bar-charts}
\end{figure}

Tab.~\ref{tab:story-stats} provides output statistics on the longest and shortest texts in our corpus. 
There were no significant performance differences based on story length, or between well- and lesser-known texts.
However, the \textbf{length of scenes} extracted by our pipeline differed by story type (Fig.~\ref{fig:bar-charts}A left).
LLM-written stories had shorter scenes (mean: 33.3 lines) than human plays (mean: 124.1) and non-plays (mean: 52.4).

\textbf{Quote accuracy} -- the percentage of real (\ie non-hallucinated) quotes extracted by the LLM -- also varied (Fig.~\ref{fig:bar-charts}A right). 
Plays scored the highest (mean: 0.97), followed by LLM-generated stories (mean: 0.90) and non-plays (mean: 0.85).
These results underscore the importance of our correction loops;
without them, the LLM returns a non-trivial number of hallucinated quotes.
Similarly, accuracy was higher when finding quotes associated with \textit{themes} (mean: 0.94) compared to \textit{characters} (mean: 0.79),
likely because character attribution requires subtle contextual clues when names are not explicitly mentioned~\cite{michel2024improving}.
This also makes sense given that LLMs were best at finding quotes in plays, which are largely composed of labeled dialogue.

\subsubsection{Study Guide Analysis}\label{sec:baseline_analysis}
As a baseline comparison, we examined SparkNotes and LitCharts study guides, which contain human-written analyses of well-known works.
In particular, we compared our extracted characters, themes, and key events to the lists and chapter summaries provided by these guides.
We analyzed \numguides stories: \textit{The Great Gatsby}, \textit{Alice in Wonderland}, \textit{Romeo and Juliet}, \textit{The Odyssey}, \textit{Pygmalion}, and \textit{Don Quixote}. 

\begin{table}[t]
\centering
\footnotesize
\caption{List of all \totalcount stories we processed with our LLM analysis pipeline. We include \famous well-known, \lessknown lesser-known, and \llmstory LLM-generated stories.}
\label{tab:stories}
\begin{tabularx}{\linewidth}{@{}X@{}}
\storybox{
\textbf{Novels ($n={\novelcount}$)}: \famous Alice in Wonderland, Anne of Green Gables, Candide, Don Quixote, Emma, Frankenstein, Great Expectations, Jane Eyre, Little Women, Pride and Prejudice, Tale of Genji, The Great Gatsby, The Metamorphosis, The Trial, The Wizard of Oz, Ulysses, War and Peace, \lessknown Under the Mendips, Dream of the Red Chamber, The Marrow of Tradition, The Tenant of Wildfell Hall

\vspace{3pt}

\textbf{Plays ($n={\playcount}$)}: \famous Hamlet, Pygmalion, Romeo and Juliet, \lessknown Faust, The School for Scandal

\vspace{3pt}

\textbf{Poems ($n={\poemcount}$)}: \famous The Iliad, The Odyssey

\vspace{3pt}

\textbf{Non-fiction ($n={\othercount}$)}: \lessknown Queen Victoria, The Art of War

\vspace{3pt}

\textbf{LLM-generated novels ($n={\llmcount}$)}: \llmstory Starlight Refugees, The Bookstore of Forgotten Dreams, The Color Thief, Threads of the Infinite, Time-Looped Detective, Whispers of the Tea Route
}
\end{tabularx}
\end{table}

\begin{table}[]
\centering
\footnotesize
\caption{Output statistics from our pipeline on the longest and shortest human and LLM-generated stories (\girl = characters, \house = locations, \theme = themes, \chat = quotes). Some story titles are abbreviated for space.}
\label{tab:story-stats}
\begin{tabularx}{\linewidth}{@{}p{2.15cm}p{0.7cm}p{1cm}p{0.7cm}p{0.25cm}p{0.25cm}p{0.25cm}p{0.25cm}@{}}
\textbf{Story} & \textbf{Lines} & \textbf{Chapters} & \textbf{Scenes} & \girl  & \house & \theme & \chat             \\ \midrule
\famous Ulysses & 25435 & 18 & 190 & 271 & 138 & 274 & 583 \\
\famous Metamorphosis & 1752 & 3 & 24 & 10 & 5 & 26 & 82 \\
\llmstory Whispers & 1741 & 12 & 61 & 25 & 23 & 75 & 165 \\
\llmstory Bookstore & 1388 & 12 & 41 & 14 & 7 & 42 & 103 \\
\end{tabularx}
\end{table}
\noindent \textbf{Method.}
We analyzed one chapter or act from the start, middle, and end of each text to study performance across narrative sections.
To compare chapter events, we listed key events from (1) our scene data, (2) SparkNotes summary, and (3) LitCharts summary.
We then performed a diff-style comparison to identify discrepancies (including in chronology).
Characters and themes were qualitatively matched when different names likely referred to the same entity (\eg ``the inevitability of fate'' vs. ``fate'').
We report the percentage of overlapping characters, themes, and events between each pair of sources, and across all three.

\begin{figure}
    \centering
    \includegraphics[alt={Three Euler diagrams showing the overlap of characters, themes, and events extracted by our pipeline, SparkNotes, and LitCharts.}, width=\linewidth]{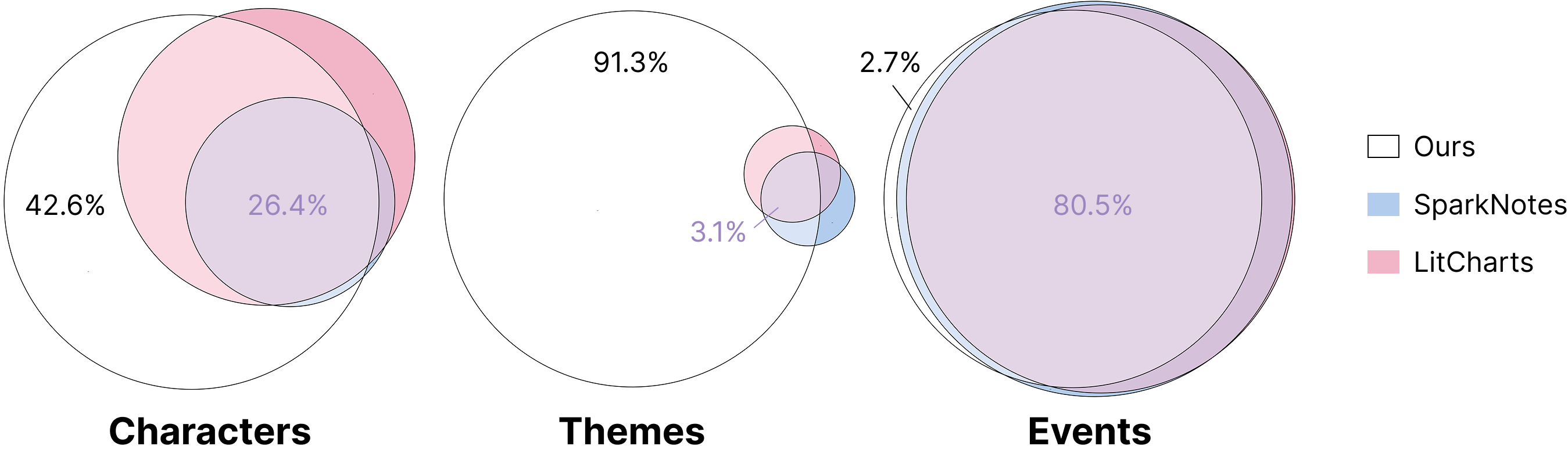}
    \caption{Visualizing the overlap of key \textit{characters}, \textit{themes}, and \textit{events} extracted by our pipeline vs. literary study guides for $n={\numguides}$ stories.}
    \label{fig:overlap}
\end{figure}
\noindent \textbf{Results.}
No major differences emerged across story sections.
On average, our pipeline extracted 94.3\% of \textbf{characters} from SparkNotes and 83.3\% from LitCharts.
The guides shared a 44.3\% overlap, and all three sources had a 26.4\% overlap (Fig.~\ref{fig:overlap}).
Most characters we missed were minor or non-speaking, \eg in \textit{Romeo and Juliet}, both guides listed Rosaline (who does not speak or appear), while the LLM did not.

For \textbf{themes}, we shared a mean 73.8\% overlap with SparkNotes and 72.6\% overlap with LitCharts.
The LLM tended to miss more complex themes (\eg ``Incompatible Systems of Morality''), or those related to language aspects (\eg ``Language and Wordplay'') and broader context (\eg ``The Roaring Twenties'').
The low overlap between SparkNotes and LitCharts (36.0\%) -- and minimal 3-way overlap (3.1\%) -- also highlights the subjectivity of identifying literary themes.

42.6\% of total characters and 91.3\% of themes were only detected by the LLM.
In \textit{Alice in Wonderland}, for instance, we found 27 characters beyond SparkNotes' list ($n=7$), and 15 beyond LitCharts' ($n=19$), \eg the Eaglet and Baby. 
Similarly, while these guides focused on analyzing 3-5 key themes, our pipeline often picked up 30+ (sometimes even 100s -- Tab.~\ref{tab:story-stats}).
Additional themes the LLM found in \textit{Romeo and Juliet} include ``Existentialism,'' ``Political Manipulation,'' and ``Inaction and Reflection.''
However, some of these characters and themes may be too minor or granular to provide user value (see Sec.~\ref{sec:llm_limitations}).

We identified 90.4\% of the same \textbf{events} as SparkNotes and 90.1\% as LitCharts with 100\% accurate chronology.
The study guides overlapped 93.6\% (3-way overlap: 80.5\%).
Our data is more structured and concise than prose summaries, but the LLM still captured most major scenes.
2.7\% of events were only extracted by the LLM (\eg when Nick confronts Meyer Wolfsheim after Gatsby's death in \textit{The Great Gatsby}).

\subsubsection{Scene Boundary Analysis}\label{sec:scene_analysis}
While discrepancies in scene length may be expected due to inherent structural differences between stories (\eg our LLM-generated texts are shorter and less complex than human writing), we wanted to learn more about the LLM's understanding of a literary scene.

\noindent \textbf{Method.} 
To do this, we annotated a total of $n=3796$ scene divisions across all \totalcount stories (excluding the first scene in each chapter).
Each boundary was labeled by examining the explanation provided by the LLM when starting a new scene and grouping these thematically.

\noindent \textbf{Results.} 
As shown in Fig.~\ref{fig:bar-charts}B, our pipeline extracted 5 main scene division types, meaning the LLM still deviates from its prompt (\ie location).
However, the most common type was \textit{location change}, as expected, making up 72.7\% of all scene boundaries.
For LLM-generated stories in particular, 85.9\% of boundaries were location-related.

The next most frequent scene division for LLM-generated stories and plays was \textit{character change}, where characters enter or exit the scene, making up 10.6\% of all scene transitions (or 15.5\% for plays).
10.5\% of scene boundaries occurred when the text's \textit{focus shifted} (\eg ``The conversation shifts to their future political strategies''),
which were the second most common division type for non-plays (11.1\%).
\textit{Character action} (\eg ``Emma formulates a plan for Harriet's future.'') and \textit{time change} transitions (\eg ``K. returns to the office the next day'') were the least frequent, making up 3.9\% and 2.2\% of all annotated scenes.

Our results reinforce how narrative scene segmentation is a challenging task for AI~\cite{zehe2021detecting}, and show that like humans, LLMs' perceptions of a scene may vary based on genre.
C2 found it interesting how 
\textit{``the LLM allows us to question seemingly minor things we take for granted. Like what exactly is a scene? Is it related to setting, character, both?''}

%% file: sections/05-design.tex
\begin{figure}[t]
    \centering
    \includegraphics[alt={Snippets of our Pride and Prejudice ribbon visualization, where the y-axis plots (A) each character in their own horizontal lane, or (B) the relative importance of each character over time.},width=\linewidth]{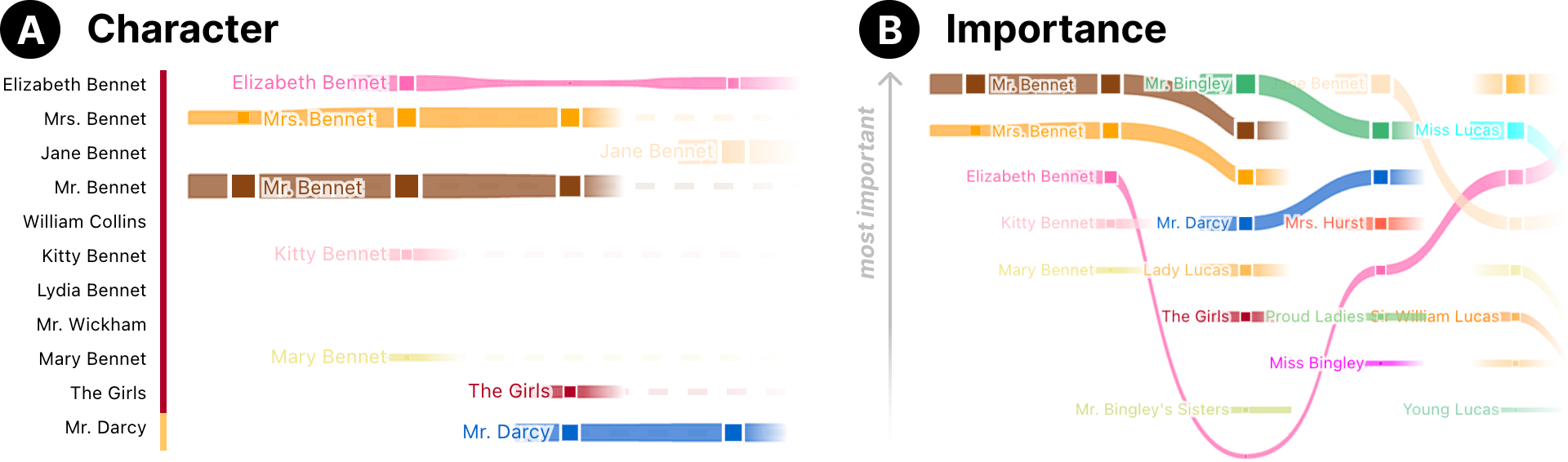}
    \caption{ \textbf{(A)} Static \textit{character} or \textbf{(B)} dynamic character \textit{importance} y-axis.}
    \label{fig:y-axis}
\end{figure}
\begin{figure*}
    \centering
    \includegraphics[alt={Screenshot showing the (A) Detail Overlay view that appears when clicking on a chapter and the (B) popup with additional scene details that appears when hovering over the main plot -- both using data from Pride and Prejudice.},width=\linewidth]{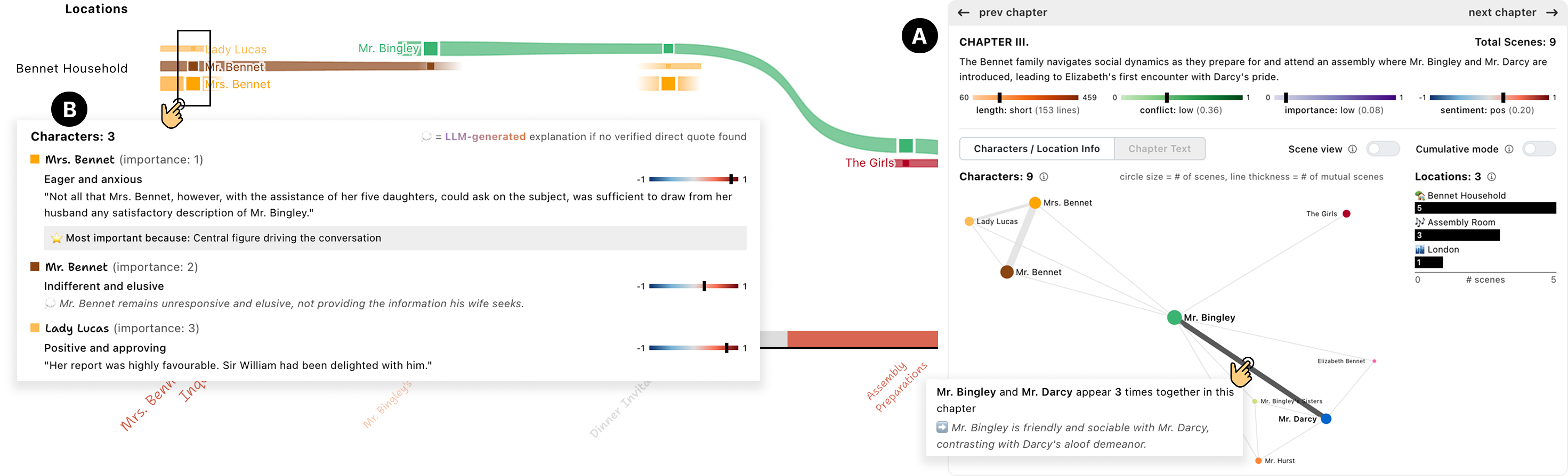}
    \caption{\textbf{(A)} When the user clicks on a chapter, the \textit{Detail Overlay} opens on the right, revealing additional details and \point a network visualization of character interactions.
    \textbf{(B)} Users can also toggle the main plot to view scenes from this chapter and \point hover for character details.}
    \label{fig:overlay}
\end{figure*}
\section{Story Ribbons Design}
Guided by our design tasks (Sec.~\ref{sec:tasks}) and \textbf{\reflink{rq2}{RQ2}}, we designed and implemented \system, an interactive tool for literary analysis.
\system visualizes narrative insights from our
LLM-powered analysis pipeline (Sec.~\ref{sec:pipeline}), allowing users to explore customizable character and theme trajectories for \totalcount stories (Tab.~\ref{tab:stories}).
While most story visualizations track interactions at a fixed time scale (\eg chapter \textit{or} scene)~\cite{kim2017visualizing,liu2013storyflow,tang2020plotthread,tang2018istoryline,watson2019storyprint}, we aim to provide a more comprehensive overview of story structure and evolution through visualizing detailed narrative information at multiple levels (\eg chapter \textit{and} scene).

\subsection{Technique}
We adapt the original storyline visualization technique~\cite{kim2017visualizing,munroe2009xkcd,tang2020plotthread}, where each character is represented by a trajectory of points -- \ie a ``ribbon'' (see Fig.~\ref{fig:teaser}).
Each point in a character trajectory denotes a particular moment in time (\eg a scene) where they are present.
Contiguous regions of points are connected by a curve, and gaps denote character absences.
Thus, drawing multiple trajectories in parallel can elicit the ``shape'' of a story.
One addition we make to this technique is using weighted paths to encode \textbf{importance}, where the thickness of a ribbon at any point reflects the character's significance in that scene.

We chose to base our work off this technique due to its established utility for visualizing narratives and flexibility for extension.
However, our experts also suggested including other visualizations, such as a character network (Sec.~\ref{sec:detail_overlay}), and so a core strength of \system is its support for switching between multiple views of a story.

\subsection{Key Interaction Motifs}
Our system includes three key LLM-powered interaction motifs:

\noindent \purplespark \textbf{Explanations on demand.}
When showing scholars (\textbf{C1-3}) early prototypes of \system, several asked questions like: \textit{``How is the LLM [deciding] what is a scene, and what sentiment and importance to assign to a character?''} (C1).
Thus, we provide ``explanations on demand,'' where the user can view explanations for AI-generated data by interacting with the corresponding UI components -- similar to Ben Shneiderman's famous ``details on demand'' mantra for interactive visualization systems~\cite{shneiderman2003eyes}.
These explanations are designed to empower users to interrogate and verify the LLM's reasoning toward increased transparency and trust \textbf{[\reflink{t4}{T4}]}.

\noindent \pinkspark \textbf{Natural language dimensions.}
\system also allows users to add new visualization dimensions through custom natural language prompts.
For example, users can ask the LLM to rank the characters in each scene by an attribute like ``sense of duty,'' or assign colors to characters based on how ``evil'' each of them is.
In this way, users are not restricted to a fixed set of traits and can shape story exploration around their own interpretative goals \textbf{[\reflink{t3}{T3}]}. 

\noindent \orangespark \textbf{Natural language queries.}
As with any new, complex visualization, users may not know where to begin or which parts are most relevant to their interests.
Thus, we include several ``ask LLM'' widgets throughout the tool to support exploration through natural language queries.
With these widgets, the user can ask a question to receive LLM guidance in navigating our visualizations and understanding story insights more deeply \textbf{[\reflink{t4}{T4}]}.
For example, if the user asks, ``Where is the theme of social class most prominent?'', the LLM will guide them to the corresponding 
chapter and segment of the visualization.

\subsection{Interface}
\system consists of three views for exploring our storyline visualizations: \textit{Story Overview}, \textit{Detail Overlay}, and \textit{Settings Sidebar}.

\subsubsection{Story Overview}\label{sec:story_overview}
The \textbf{Story Overview} contains our main ribbon visualization (Fig.~\ref{fig:teaser}).
In the top left corner, users can view story metadata.
To the right, a menu bar provides visualization controls, \eg the ``Character view'' toggle switches between visualizing \textit{characters} and \textit{themes} \textbf{[\reflink{t2}{T2}]}.

Below, the ribbon plot maps time on the x-axis, segmented by \textit{chapter} or \textit{scene} (via the ``Chapter view'' toggle) \textbf{[\reflink{t1}{T1}]}.
The color of each chapter label, and the corresponding band, encodes \textit{sentiment} (adjustable in Settings).
By default, the y-axis encodes \textit{location} in order of chronological appearance in the text.
We use a consistent y-axis to improve readability and preserve context about character interactions~\cite{arendt2017matters}; this diverges from traditional storyline visualizations where y-coordinates encode interaction via proximity~\cite{munroe2009xkcd,tang2020plotthread}.
Users can switch the y-axis to track other narrative aspects as well \textbf{[\reflink{t2}{T2}]}:
\begin{itemize}
    \item \textit{Character}: plots each character ribbon in its own horizontal lane, organized by group, to help identify co-occurrences (Fig.~\ref{fig:y-axis}A).
    \item \textit{Importance}: plots ranked character importance over time, with more prominent characters at the top (Fig.~\ref{fig:y-axis}B)\footnote{We plot rankings instead of raw importance scores (0-1) because the LLM often assigns similar ratings to characters, making interpretation more difficult.}.
    \item \textit{Sentiment}: plots character sentiment over time, with \textcolor{DarkBlue}{-1: negative} at the bottom and \textcolor{DarkRed}{1: positive} at the top.
\end{itemize}
We also support ``Rank by trait,'' a \customize{natural language dimension} that adds a custom y-axis for users to explore by ranking characters or themes in each scene by a specific trait (\eg inner conflict) \textbf{[\reflink{t3}{T3}]}. 

\subsubsection{Detail Overlay}\label{sec:detail_overlay}
After gaining a high-level overview, users may want to explore the narrative in more fine-grained detail \textbf{[\reflink{t1}{T1}]}.
To do this, users can invoke the \textbf{Detail Overlay} by (1) clicking on a chapter in the main plot, or (2) using the ``Ask LLM'' button near the x-axis (Fig.~\ref{fig:teaser}).
(2) opens a prompt box, where users can ask a question about the story (\eg ``When does Elizabeth reject Darcy?'').
The LLM identifies the most relevant chapter to the user's \explore{natural language query} and provides an explanation \textbf{[\reflink{t4}{T4}]}, directing them to the corresponding overlay.

In this view, users can explore the selected chapter in depth (Fig.~\ref{fig:overlay}A).
At the top, we show the chapter summary and ratings (length = normalized number of lines) \textbf{[\reflink{t2}{T2}]}.
Below, there is an interactive network visualizing interactions (links) between characters (nodes) in this chapter (or the story \textit{through} this chapter if ``Cumulative mode'' is on)~\cite{min2019modeling,john2019visual,labatut2019extraction}.
The size of each character node encodes their chapter importance, based on the number of scenes they appear in.
Edge thickness encodes character co-occurrences. 
Users can hover on a character or interaction for an \explain{explanation} about their role or relationship in this chapter \textbf{[\reflink{t4}{T4}]}.
On the right, there is a bar chart with chapter locations.

Above the network, users can toggle to ``Scene view'' (instead of ``Chapter view'') to visualize only scenes \textit{within this chapter} in the main ribbon plot \textbf{[\reflink{t1}{T1}]}.
Hovering on a scene will open a similar overlay with a list of characters that are present in the scene, ranked by importance, as in Fig.~\ref{fig:overlay}B.
For each character, we show the LLM's description of their emotions and a corresponding quote \textbf{[\reflink{t5}{T5}]}. 
Here, the LLM did not find a direct quote for Mr. Bennet, so an explanation is displayed instead.
Users can also see the LLM's \explain{explanation} for why it chose the top character as the most important in this scene \textbf{[\reflink{t4}{T4}]}.

\begin{figure}
    \centering
    \includegraphics[alt={Screenshot showing our system's view of the chapter text from Pride and Prejudice, along with corresponding interactions.},width=\linewidth]{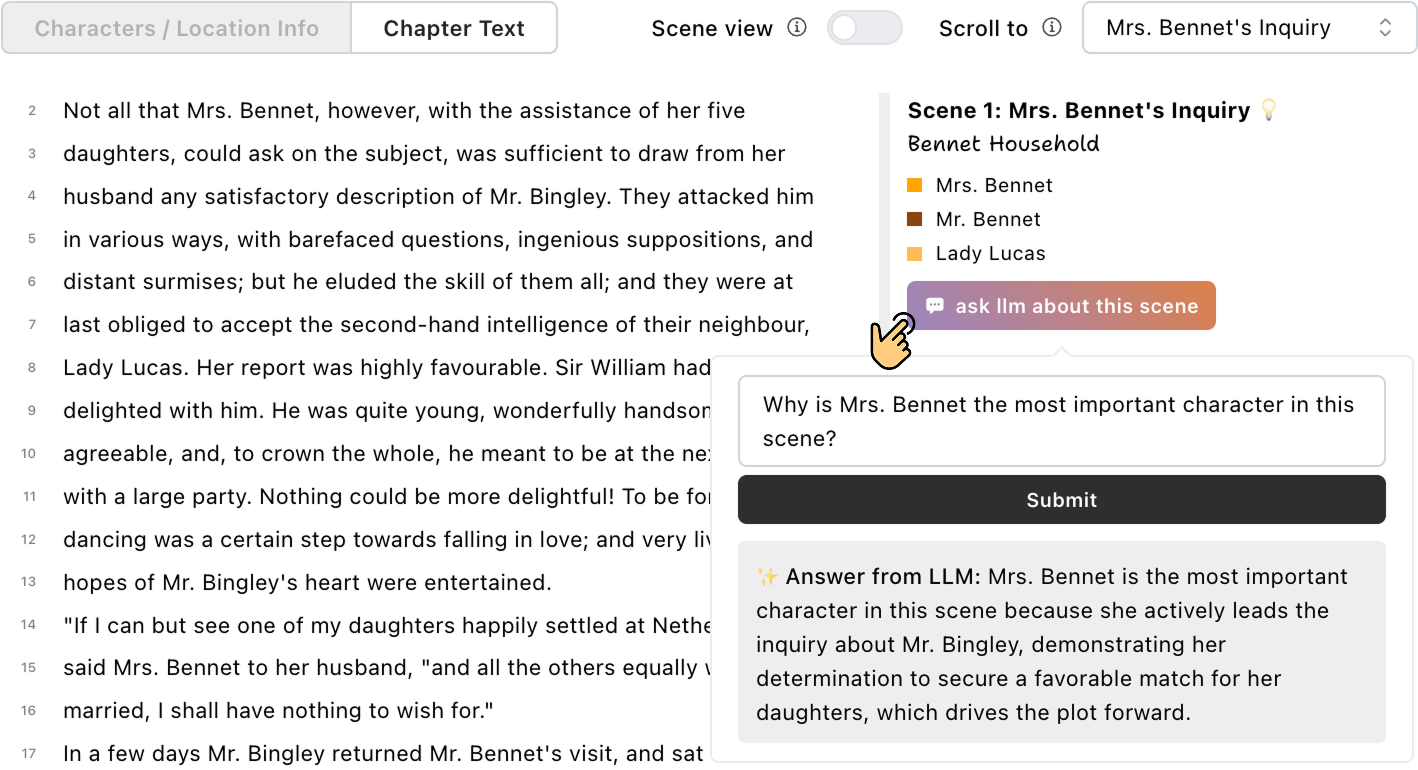}
    \caption{Viewing raw chapter text inside of the \textit{Detail Overlay}. \point Here, users can ask the LLM a question about the current scene or chapter.}
    \label{fig:chapter-text}
\end{figure}

To view the ``Chapter Text,'' users can click the corresponding button \textbf{[\reflink{t5}{T5}]} (Fig.~\ref{fig:chapter-text}).
Our visualizations automatically scroll as you read and show the corresponding scenes on the right.
Hovering on the \light icon next to a scene title displays the LLM's \explain{explanation} for starting a new scene \textbf{[\reflink{t4}{T4}]}.
Users can ``ask LLM about this scene'' (or chapter), which uses the story text to promote deeper exploration of the narrative or the LLM's decision-making process through \explore{natural language queries}.

\subsubsection{Settings Sidebar}\label{sec:sidebar}
To further customize the ribbon plot, users can open the \textbf{Settings Sidebar} via the button in the top right corner of Story Overview (Fig.~\ref{fig:teaser}).

\begin{figure}
    \centering
    \includegraphics[alt={Screenshot of our Settings Sidebar, which contains various options for customizing (A) character and (B) chapter views. (C) Shows Elizabeth Bennet's ribbon being highlighted in the main plot, and her overlay box above with a quote and explanation.},width=\linewidth]{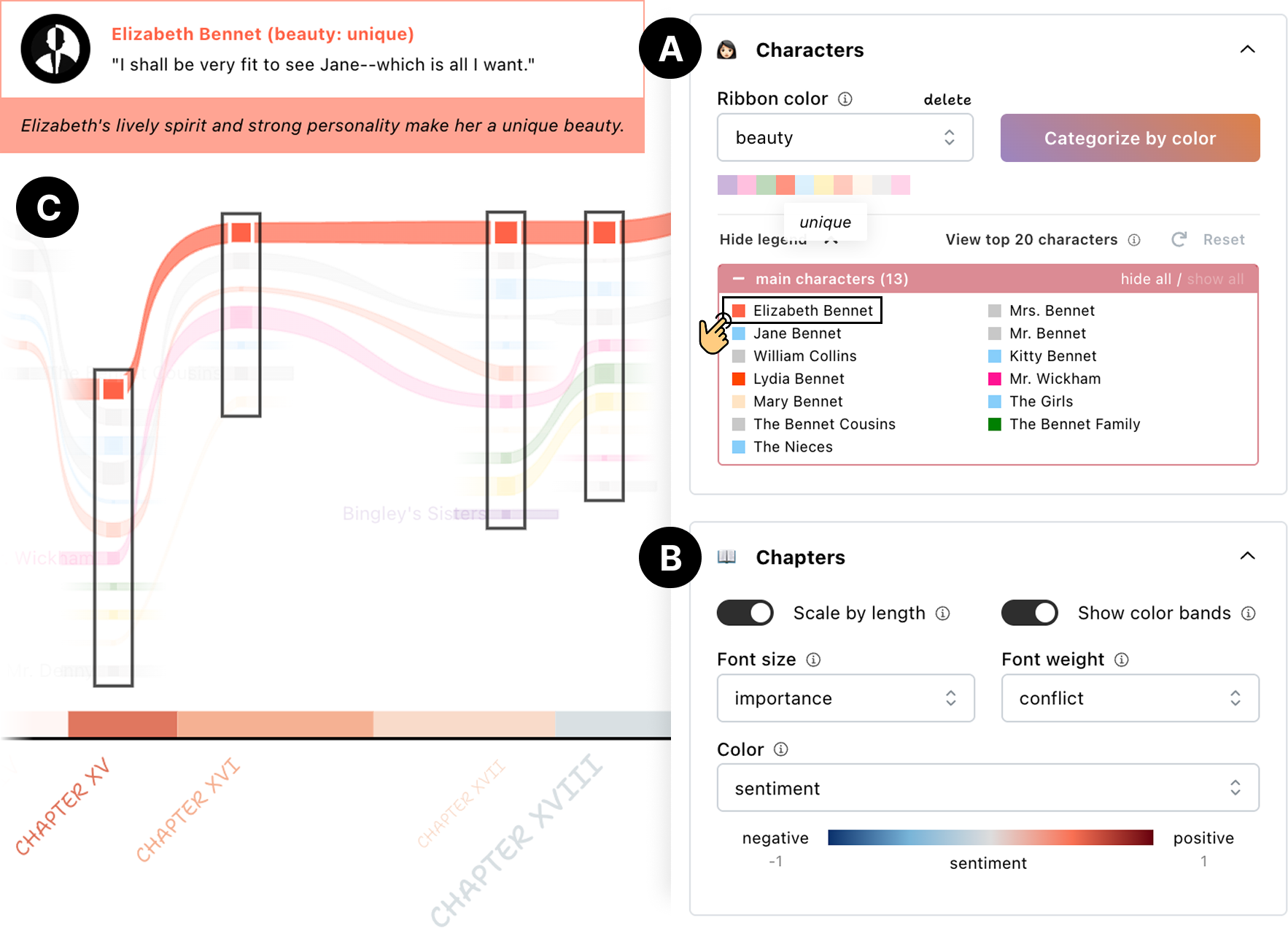}
    \caption{Users can customize \textbf{(A)} character ribbons and \textbf{(B)} chapter labels in the \textit{Settings Sidebar}.
    \textbf{(C)} \point
    Hovering on a character highlights their ribbon in the main plot, along with the corresponding overlay.}
    \label{fig:sidebar}
\end{figure}

In the ``Characters'' settings, users can change the ribbon colors (Fig.~\ref{fig:sidebar}A).
By default, we use the \textit{LLM-assigned} colors, but ribbons can also encode character \textit{group}, \textit{importance}, or \textit{sentiment} \textbf{[\reflink{t2}{T2}]}.
Group colors are chosen from a discrete color scale, while importance uses a continuous, sequential scale and sentiment uses a continuous, diverging scale between \textcolor{DarkBlue}{blues (negative)} and \textcolor{DarkRed}{reds (positive)}.
Users can also ``Categorize by color'', a \customize{natural language dimension} that adds custom color palettes based on an attribute (\eg wealth or age) \textbf{[\reflink{t3}{T3}]}.
The LLM assigns each character or theme a value of the specified attribute (\eg lower, middle, or upper class),\footnote{
We use discrete values for easier attribute encoding and user interpretation.} along with a corresponding color.

Under the color dropdown, there is a filterable character legend organized into groups.
Here, users can find characters or subgroups of interest and highlight/hide their corresponding ribbons.
Upon hovering on a character in the legend, or a ribbon in the main plot, a popup opens with an LLM-selected quote about the character and \explain{explanation} for their current color encoding (\eg why Elizabeth is a \textcolor{CoralRed}{``unique''} beauty) \textbf{[\reflink{t4}{T4}]} (Fig.~\ref{fig:sidebar}C). 
Hovering on a location brings up a similar popup.

Below, in the ``Chapters'' settings, users can customize the chapter labels and corresponding color bands (Fig.~\ref{fig:sidebar}B).
Label color, size, and weight can be used to encode information about chapter \textit{importance}, \textit{sentiment}, \textit{conflict}, \textit{length} (normalized), and \textit{number of characters} (normalized) \textbf{[\reflink{t2}{T2}]}.
Like the character ribbons, continuous, sequential color scales are used to depict each characteristic.
Users can toggle ``Scale by length'' to scale the width of each color band to reflect true chapter length; by default, the x-axis is broken into equal segments.

\subsection{Implementation}\label{sec:sys_implement}
\system is a full-stack web application with a Python/Flask backend that communicates with a React/Typescript frontend.
Our main ribbon plot is a custom SVG visualization constructed with Bézier curves~\cite{mortenson1999mathematics}.
The character network is implemented using \href{https://d3js.org/}{D3.js}.

Our backend uses \href{https://www.langchain.com/}{Langchain} and \href{https://platform.openai.com/docs/models/gpt-4o-mini}{\texttt{gpt-4o-mini}} to power all on-the-fly features (\eg ``Ask LLM'', ``Rank by trait'') and most story analysis steps due to overall high output quality as well as speed and cost efficiency.
On-the-fly operations take the user's query and our extracted story data as input.
Our LLM prompts and example user queries are included in the supplementary materials.

For correction loops, we use \href{https://www.anthropic.com/news/claude-3-5-sonnet}{\texttt{claude-3-5-sonnet}} as our second LLM to group duplicate story elements, as Claude outperformed GPT on this task during pilot studies. 
The idea of leveraging multiple LLMs based on their unique strengths is inspired by work such as~\cite{venkatraman2024collabstory}.
We parallelize several pipeline steps (\eg scene splitting \& analysis) to reduce latency.
Running our pipeline on \textit{The Great Gatsby} (4710 lines) takes $2.2$ minutes, while \textit{Little Women} (16680 lines) takes $3.9$ minutes.

\subsection{Usage Example: \textit{Pride and Prejudice}}

This design was heavily informed by collaboration with our scholar co-authors. 
To illustrate how they used the tool, we provide a brief vignette from an exploration of \textit{Pride and Prejudice.}
While ranking characters by \textbf{importance}, C2 was interested in the LLM's characterization of Mr. Bennet as the most important character at the beginning of the novel: \textit{``That surprised me because it's very instinctive to read Mr. Bennet as more of a side character''} (Fig.~\ref{fig:y-axis}b).
Upon reflection, C2 said, \textit{``Now I do see how Mr. Bennet is very important. He's in the middle of the scene and everyone is talking to him... He's the one who arranges for his daughters to go to Mr. Darcy and Bingley's house.''} 

Next, C2 visualized the LLM's concept of \textbf{beauty} in the novel by adding a custom color palette, and was surprised that instead of a scale from ugly to pretty, the LLM created different categories of beauty (\eg natural, classic, exotic).
C2 was intrigued that both Elizabeth and Charlotte were classified as  ``unique'' beauties \textit{``because in the novel, Charlotte is called downright ugly. And she's Elizabeth's best friend, so she's really sad that she's not as pretty''} (Fig.~\ref{fig:sidebar}C).
Ultimately, C2 thought \textit{``this is great. The visualization makes you think about beauty in a different way... and it's not that the LLM got it wrong, but it's making us see the characters and story in a different way.''}

%% file: sections/06-eval.tex
\section{\system Evaluation}\label{sec:evaluation}

We evaluated our tool through a user study and expert interviews, synthesizing findings with feedback from our original scholars (\textbf{C1-3}).

\subsection{User Study}\label{sec:user_study}
Our user study involved asking participants to explore a story of their choice using \system and its interactive visualizations.

\noindent \textbf{Participants.} 
We recruited a total of \numparticipants undergraduates, graduate students, and working professionals who are strongly interested in literature (\textbf{P1-16}).
\numexperts participants self-identified as \textit{expert} literary analysts (\eg English major who regularly analyzes literature), \numintermediate as \textit{intermediate} analysts (\eg took multiple English classes and familiar with literary analysis), and \numnovices as \textit{novice} analysts (\eg avid reader but little to no formal literary training). 
12 out of our 16 participants had never seen a narrative visualization before.

\noindent \textbf{Procedure.} 
Prior to the study, we asked each participant to select a story they are familiar with. 
Our goal was to make the task more personalized and ensure that users could meaningfully assess the LLM's insights.
In total, participants examined 11 unique literary works.

We started each 1-hour session by asking participants a few pre-task questions about their experience with literary analysis and visualizing stories.
This was followed by a walkthrough of our interface and visual encodings using their chosen story.
Next, participants had $\sim$30 minutes to dive analytically into their story by interacting with \system and its features.
Participants were also asked to think aloud.
The study concluded with a post-task interview to gain deeper insights into each participant's experience with \system, how they perceived the LLM-generated data, and other feedback about our visualizations.

\subsection{Expert Interviews}\label{sec:exp_feedback}
We conducted semi-structured interviews with \numscholars new literary scholars to elicit feedback about \system from a scholarly perspective.
These scholars included an English Ph.D. student and two English professors (\textbf{S1-3}).
Each interview followed a similar structure as the user studies but were shorter ($\sim$30 minutes) and more focused on collaboratively exploring the system and ideating potential use cases.

\begin{figure*}
    \centering
    \includegraphics[alt={Screenshot of our visualization of \textit{The Metamophosis,} to compare (A) the default view with locations along the y-axis with (B) a custom y-axis which ranks characters by ``hope,'' and colors their ribbons by ``sense of duty.'' },width=\linewidth]{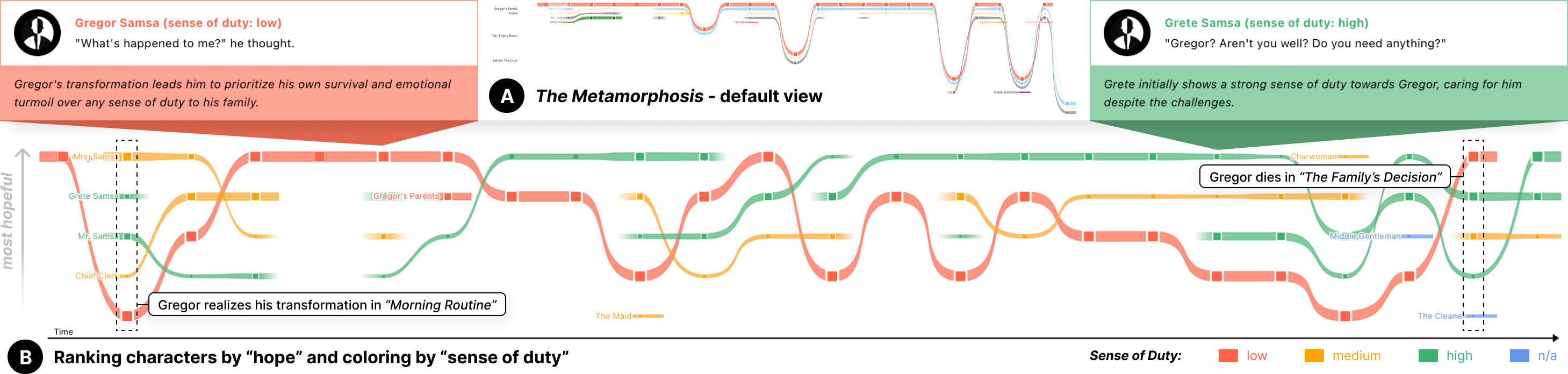}
    \caption{\system visualization of \textit{scenes} in \textit{The Metamorphosis}. \textbf{(A)} The default view with locations along the y-axis.
    \textbf{(B)} After ranking characters by ``hope'' and coloring the ribbons by ``sense of duty.'' Corresponding color explanations for Gregor and Grete are included.}
    \label{fig:metamorphosis}
\end{figure*}
\section{Results}\label{sec:results}

Overall, participants enjoyed using \system to explore stories,
noting that our system provided a powerful, intuitive way for getting a \textit{``bird's eye view''} (P6) and understanding \textit{``how a story arc develops over time''} (P3).
Users like P4 and P14 found our tool useful for revisiting \textit{``specific scenes [in] familiar stories''} and \textit{``keeping track of everything -- especially for a complex story like The Odyssey''} ($n=12$).
Our visualizations often reminded users about story elements as well: \eg \textit{``I'd forgotten there was a king of hearts!''} (P1, \textit{Alice in Wonderland}) or \textit{``I didn't realize Nick is in every chapter''} (P8, \textit{The Great Gatsby}).

\subsection{Case Study: \textit{The Metamorphosis}}\label{sec:example_findings}

To illustrate \system's utility, we share how P4 used our tool to gain new insights about \textit{The Metamorphosis} by Franz Kafka.
P4 began exploring the story by viewing our \textbf{scene breakdown} along the x-axis: \textit{``Oh, Gregor's death [is] called The Family's Decision. That's an interesting reframing because they do just decide to let him die and Gregor doesn't have much agency over anything during the story''} (Fig.~\ref{fig:metamorphosis}). 
They added, \textit{``If you compare this to SparkNotes where it's more like a description of what happens, this has a lot more sauce. It's trying to say something about the scene, like why is it there? And that's central to the point of analysis, which really fascinates me.''}

Next, P4 changed the y-axis to \textbf{hope} to trace Gregor's ups and downs: \textit{``That seems accurate. First he's shocked about the situation and then sort of accepts things. But then, his family has to get jobs and he feels bad. Then he feels better because he's like my family cares about me. But then he's like they threw an apple at me. And now I'm okay with dying.''}
P4 also found it \textit{``funny that the sentiment''} along the x-axis shows \textit{``they're feeling relieved and happier after Gregor dies''} and wondered if the LLM understood the ending's \textit{``dramatic irony.''}

However, P4 noted that the LLM seemed to miss the nuance in Grete's character.
When ranking characters by \textbf{importance}, Grete was often near the bottom: \textit{``Maybe she's more of a minor character if you count number of appearances, but [I think] she's actually one of the most important people, because of the themes about family and alienation [and] Grete's betrayal of Gregor is such an important scene.''}
Coloring by \textbf{sense of duty}, P4 said \textit{``it's interesting that [the LLM] captures Grete's caring nature towards Gregor''} in classifying her as ``high.''
They also disagreed with the LLM labeling Gregor as ``low'': \textit{``It's true Gregor's being alienated from his human self, but the primary source of his oppression is feeling like he's a burden to his family.''}

\subsection{User Feedback}\label{sec:user_feedback}
Users appreciated \system' high degree of \textbf{customization} ($n=12$): 
\textit{``The customizability is very cool. You're practically able to do anything''} (P12) and can \textit{``adapt it to what [you] want to see''} (P3).
S1 said it encouraged them to explore new story aspects: \textit{``I don't look at sentiment analysis much, but I really enjoyed seeing the sentiment view and how [the LLM] summed it up with color.''}
However, 6 users noted the visualization and customization options could be \textit{``too much at times''} (P6), especially for longer stories with many characters, and they \textit{``felt a bit overwhelmed by all the things happening at once''} (P3), suggesting opportunities to better manage cognitive overload.

Participants also emphasized that story \textbf{familiarity} makes the visual analysis process more meaningful ($n=8$): 
\textit{``I'd probably only use this tool for books that I've read because then I kind of know what to look for. If I haven't read a book, it'll take my joy away [in] discovering themes and analyses myself''} (P13).
Similarly, P9 thought \system \textit{``would be most helpful for [people] who are familiar with the book [and] want to explore new questions or interpretations.''}
C1 and P12 had concerns about people \textit{``taking [the LLM's] words as the truth and trusting it too much''} if they have not read the story before.

\subsubsection{Most \& least useful features}
14 participants enjoyed using \textbf{``Categorize by color''} as a \textit{``fun entryway into analysis''} (P4) that \textit{``lets me consider hypotheses that I might not have had otherwise''} (P2), visualizing attributes like authority, loyalty, and evilness.
Some explored applying modern concepts to classic works (\eg P11 colored \textit{Pride and Prejudice} characters by \href{https://www.myersbriggs.org/my-mbti-personality-type/myers-briggs-overview/}{MBTIs}.)
Similarly, 12 participants found \textbf{``Rank by trait''} useful for gaining \textit{``different views [and] insights''} (P5), especially in \textit{``how characters embody a theme''} (S2) such as vengeance or feminism. 
C2 also tried ranking characters by language traits such as humor or wordplay.

9 users described \textbf{``ask LLM''} as a \textit{``more precise Ctrl+F''} that can \textit{``focus me on a chapter based on my interest''} (P3), which is helpful \textit{''especially [for] sifting through a huge book''} (P9).
4 participants used it to locate key events: \eg \textit{``My favorite part is when Mr. Rochester proposed to Jane Eyre. [I'm] curious what I can see from the visualization about that''} (P13).
Users also enjoyed reading LLM-extracted quotes and explanations ($n=8$), tracking and filtering characters in the legend ($n=4$), and viewing story text to verify AI insights ($n=3$).
P2 and P11 thought the x-axis color bands were useful in \textit{``seeing how the LLM chopped up the book [and] observing trends like conflict.''}

10 participants said the \textbf{character networks} were their least favorite feature.
While some like P5 saw them as \textit{``a cool new way to interpret the text''} ($n=5$), others found it \textit{``confusing''} when \textit{``it had many nodes [and] everything was connected''} (P1).
P9 noted that \textit{``the number of interactions [is] not as important as the weight or nature of those interactions''} (P4).
Users were also split on the utility of visualizing \textbf{locations}.
P9 and P14 enjoyed seeing \textit{``the physical journey''} of characters and themes, and P12 valued this view as someone with aphantasia who \textit{``doesn't have mental imagery.''}
However, 3 users found this y-axis less informative for stories with fewer settings and 2 thought it \textit{``could be misleading if it's not ordered by proximity''} (S3).

\subsection{Limitations of LLMs}\label{sec:llm_limitations}
User also discovered several interesting behaviors and limitations of LLMs while interacting with \system.

\subsubsection{Context and granularity challenges}
We observed in Sec.~\ref{sec:pipeline_design} that LLMs may refer to the same character or location by different names.
Our correction loops aim to consolidate duplicate elements; however, this task requires \textbf{context} about the current scene or entire story, which the LLM struggles with.
As P14 noticed in \textit{The Odyssey}, both Polyphemus and Cyclops were listed as characters, but \textit{``Polyphemus is the name of the Cyclops, unless I'm remembering wrong?''}
Similarly, in \textit{The Metamorphosis}, P4 was surprised that besides Mr. and Mrs. Samsa, \textit{``there's another category of Gregor's parents. It seems [the LLM] can't infer those relationships.''}

Users also observed the LLM having trouble finding the right \textbf{granularity} for analysis. 
5 participants commented that our themes felt \textit{``not super well defined''} (P15) and \textit{``maybe too granular''} (P3).
For example, in \textit{Jane Eyre}, there were many related themes like ``nature's beauty'' and ``nature's harshness,'' but P5 was \textit{``curious about [not seeing] feminism,''} a core \textit{``theme I enjoyed in the book.''}
Users like P8 wondered how the LLM defines a character, as in \textit{The Great Gatsby}, it included T.J. Eckleberg, \textit{``a symbol with a human name,''} and very minor figures like First Girl.
P6 asked how character groups were determined, noticing that \textit{Pride and Prejudice} had ``main characters'' and ``upper class,'' which could overlap (\eg Mr. Darcy).

\subsubsection{Lack of advanced analytical capabilities}
8 participants noted the LLM's \textbf{inability to surface deeper literary insights}.
For instance, P1 asked the LLM about takeaways from \textit{Alice in Wonderland} but \textit{``its answers [were] more surface level,''} \eg the absurdity of authority.
P2 and P13 wanted \textit{``the LLM [to] perform holistic analyses''} that synthesized and \textit{``weren't so bounded by chapters.''} 

Like P4 (Fig.~\ref{fig:metamorphosis}B), P7 and P15 noticed that the LLM overlooked a lot of complexity in its character analyses, \eg fixating on Amy's selfishness and Jo's kindness in \textit{Little Women}, while \textit{``Amy and Jo are both flawed characters [and] a bit selfish in their own ways.''}
4 other participants observed similar behaviors where the LLM's explanations or \textit{``quotes aren't super representative [and] don't tell me anything interesting about how the characters relate to''} a theme (P16).

\subsubsection{Impact on trust}
14 participants said unexpected LLM behaviors \textbf{did not affect} their overall experience with or trust in our visualizations. 
8 noted, \textit{``It's not surprising there were errors''} (P14) and \textit{``we should have some degree of doubt about what [LLMs] are saying''} (P12).
As P4 put it, \textit{``it's like literary criticism. You don't have to agree. It's still interesting [and] just one point of view you can gain something from.''}
P9 added, \textit{``I believe in my own authority over the LLM,''} and P11 said, \textit{``If anything, it would just make me dig deeper to see where [the LLM] is getting at.''}

However, 5 users shared concerns about the LLM's \textbf{more subjective} ratings such as importance, as \textit{``that's such an interesting literary category and a large part of the critic's job''} (S2).
P10 was apprehensive because even with our provided explanations, \textit{``I have no idea how accurate it is [and] no gauge for what they see as important.''}
Similarly, P8 said they would be cautious about trusting ``ask LLM'' for \textit{``personal or philosophical questions that require deep analysis and discussions.''}

\subsection{Use Cases \& Applications}\label{sec:use_cases}

\noindent \book \textbf{Pedagogy.}
Both participants and scholars saw immense potential for a tool like \system in pedagogical settings.
7 users described how looking at \textit{``themes and unexpected connections could [be] great for starting discussions'' or ``coming up with writing prompts for high schoolers or undergrads''} (S1).
8 participants mentioned that our tool could provide valuable essay writing support as well: \textit{``If I were writing an essay [but] didn't flag everything while reading, then I could use the tool to find quotes and text evidence that I forgot''} (P9).

\noindent \chat \textbf{Book clubs.}
Similarly, 4 users envisioned our tool facilitating discussions in book clubs.
P2 noted, \textit{``This would be fascinating to have the whole timeline [and] help us see where the interesting discussions are.''}
P7 said our visualizations \textit{``would be fun for people [to] look at things together and see if they got the same things out of a book.''}

\noindent \light \textbf{Scholarship.}
Some experts noted that in its current form, \system may not fit their specific workflows, but they ideated ways of adapting our tool to meet various scholarly goals.
4 were interested in using LLMs to incorporate multiple perspectives or \textit{``knowledge sources like SparkNotes and other literary experts''} into our visualizations to provide \textit{``meta analysis about the author [and] cultural context''} (C2, P4).
8 users wanted to add a more comparative dimension, enabling 
cross-novel analysis and ``reading at a nonhuman scale [where] you could query across a large corpus of works'' (C3, S2).
Users also asked to visualize different kinds of texts, including historical timelines (S3), nonlinear narratives (P6), and stories in other languages (S1, P9).

\noindent \theme \textbf{Writing.}
7 participants suggested extending our tool to have more of a writing or authoring focus. 
For instance, P5 wanted to modify the visualization to explore ``what if'' questions like: \textit{``Hamlet is a tragedy [but] let's say Hamlet survived. How would the story be?''}
C3 also imagined using LLMs to \textit{``regenerate stories from different perspectives [or] fill in missing gaps,''} while S3 saw an opportunity to integrate similar storyline visualizations \textit{``inside of writing tools to help writers while writing complex novels and TV shows.''}

%% file: sections/07-discussion.tex
\section{Discussion \& Future Work}
Our results offer insight into current limitations and opportunities for AI-enhanced text visualizations.

\subsection{Sparking and Drawing From Literary Conversation}

Multiple users suggested that \system could serve as a valuable \textbf{conversation catalyst}, particularly in group settings such as classrooms or book clubs (Sec.~\ref{sec:use_cases}). 
That is, the visualization could serve as a visual aid or reminder of a book, consistent with cognitive psychology research on the effectiveness of visual memory aids~\cite{spence1985memory}. 
Interestingly, several users observed that although the LLM's interpretation of narrative elements did not always align with their perspectives, that in itself could spark conversation in meaningful ways (Sec.~\ref{sec:user_feedback}).

Several participants noted the fact that our system is limited by only having access to the story text (\eg P15: \textit{``if I were to read this book in a vacuum, I'd probably miss a lot of the things the LLM did''}). P5 also reported the importance of \textit{``knowing what others have observed [and] their perspectives on how pivotal a given part is in the story.''} 
An important future direction would be to give \system access to this \textbf{wider literary conversation}, adding critical work to the base text.

\subsection{Toward More Integrated LLM-Visualization Interactions}
Users hinted at the value of tightly integrating LLM insights and their corresponding visualizations for \textit{authoring} and \textit{question-answering}.

\noindent \textbf{Visual authoring.}
After interacting with \system, many users wanted to manipulate our ribbon plots and see how the underlying stories would change (Sec.~\ref{sec:use_cases}).
Some works are starting to explore this idea of linking text and visualizations for creative storytelling.
For example, \cite{masson2024visual} introduces the idea of ``visual writing,'' where users can author stories by manipulating character traits and timelines.
TaleBrush~\cite{chung2022talebrush} and Patchview~\cite{chung2024patchview} are two other visualization tools that champion interactive story sketching and worldbuilding with LLMs.

In this way, LLM hallucinations can be leveraged for ``good,'' enabling new modes of visual authoring and literature exploration.
P2 and S3 thought our tool's visualization of scene structure could be especially valuable for writers to learn \textit{``how different authors present [their] stories''} and \textit{``what a literature-quality story looks like visually.''}

\noindent \textbf{Visual question-answering.}
Our work suggests new directions for visual question-answering, which typically focuses on answering questions about image and video data~\cite{antol2015vqa}.
We propose extending this paradigm to text data, where users interact with dynamic visualizations to explore and answer literature analysis questions.

While related to prior work on using natural language to generate or modify data visualizations~\cite{narechania2020nl4dv,vaithilingam2024dynavis}, we envision a more integrated workflow where LLMs proactively guide exploration of user queries by customizing visualizations and directing attention to relevant aspects of the text.
For example, users like P12 wanted to directly probe literary visualizations with LLMs: \textit{``I'd like to pick a section of the ribbon and [have] the LLM tell me something about that,''} going beyond our current natural language features.
P14 also suggested having a linked history view for ask LLM requests that \textit{``shows you all the questions [and] where you asked them in relation to''} a visualization.

\subsection{Scaling Analytical Power and Trust: A Tradeoff}
Another emergent theme was the tradeoff between wanting to leverage LLM-infused visualizations for more complex and large-scale tasks but having concerns about output faithfulness (Sec.~\ref{sec:llm_limitations},~\ref{sec:use_cases}).
Most participants agreed that the LLM was best at \textit{``objectively organizing things like where certain characters and themes pop up''} that are \textit{``tangible and concrete''} (P9, P10).
Users hoped to capture more nuanced analyses (\eg cross-chapter comparisons or philosophical queries), but were hesitant to trust the LLM in these cases, as \textit{``it's harder to know what it's looking for''} (P8), echoing findings from~\cite{yang2024analyzing}.

Participant sentiments highlight the larger need for greater \textbf{explainability and interpretability}, especially in fields like literature where \textit{``someone is passionate about it and wants details. So if [the LLM] misses out and hallucinates, that's a major lack in trust''} (P5).
Our participants were on average highly educated and likely more familiar with LLMs than the general public, but other users may be prone to AI overreliance.
While we provide explanations for LLM decisions, these are currently prompt-based and may not always be fully faithful to model internals -- highlighting an important area for future research.
For example, literary visualization tools could include more robust explainability measures such as confidence scores (P3)~\cite{huang2023look} or RAG-based textual evidence for LLM responses (P7, P9)~\cite{wang2024ragviz,lewis2020retrieval}.
This is particularly crucial for distant reading applications~\cite{moretti2013distant} where scholars compare works at scale, and \textit{``it would be harder to know whether to trust the LLM's interpretations, especially if I'm not familiar with everything it's showing''} (S2).

%% file: sections/08-conclusion.tex
\section{Conclusion}
\system augments traditional storyline visualization techniques with AI text-processing capabilities. 
Our system is based on a custom LLM-powered data processing pipeline, which extracts detailed semantic content from novel-length stories. 
The pipeline design uses multiple ``correction loops'' to make it resilient to AI errors. 
\system' visualizations extend the expressive power of standard storylines, with additional visual encodings to display the rich details produced by the data pipeline. 
Our tool also supports the interactive use of LLMs to produce explanations on demand and add custom visualization dimensions to personalize story exploration.

Feedback from a user study, along with conversations with literary scholars, demonstrate that \system can provide new insights about and illuminate interesting paths for further exploration of a text. 
At the same time, our findings reveal limitations in what current LLMs can achieve -- particularly in grappling with literary nuance, understanding narrative context, and resolving ambiguities. 
Even so, our results suggest that \system represents a promising tool for literary analysis, and its design strategies for the integration of AI may be helpful in creating and enhancing other visualizations.

%% file: sections/00-acknowledgements.tex
\acknowledgments{
We thank the participants in our user study and expert interviews for
their time and invaluable insights.
Additionally, we would like to thank Andrew Lee and Lucia Gordon, along with the anonymous reviewers, for their helpful feedback and suggestions in revising this paper.
We are also grateful for the support provided by the members of the Harvard Insight + Interaction Lab.

This research was supported by CY's National Science Foundation Graduate Research Fellowship under Grant No. DGE 2140743 and Kempner Institute Graduate Research Fellowship.
MW and FV received support from the Effective Ventures Foundation, Effektiv Spenden Schweiz, an Open AI Superalignment grant, and the Open Philanthropy Project.
}